\documentclass[12pt]{article}
\usepackage[utf8]{inputenc}
\usepackage{pifont}
\usepackage{amssymb}
\usepackage{amsmath}
\usepackage[french]{babel}
\usepackage[francais]{varioref}
\usepackage[dvips]{graphicx}

\usepackage{fancyhdr}
\usepackage{titlesec}
\usepackage{vmargin}

\usepackage{subfig}

\usepackage{multirow}
\usepackage{flafter}
\usepackage{eurosym} 
\usepackage{array}

\usepackage{color}
\usepackage[%
    colorlinks=true,
    linkcolor=blue,
    citecolor=magenta]{hyperref}

\usepackage{pstricks}

\usepackage{psfrag}

\makeatletter
\renewcommand{\thesection}{\@arabic\c@section}
\makeatother

\DeclareFixedFont{\chapnumfont}{OT1}{phv}{b}{n}{80pt}
\DeclareFixedFont{\chapchapfont}{OT1}{phv}{m}{it}{40pt}
\DeclareFixedFont{\chaptitfont}{OT1}{phv}{b}{n}{25pt}

\definecolor{gris}{gray}{0.75}
\titleformat{\chapter}[display]%
	{\sffamily}%
	{\filleft\chapchapfont\color{gris}\chaptertitlename\
	\\
	\vspace{12pt}
	\chapnumfont\thechapter}%
	{16pt}%
	{\filleft\chaptitfont}%
	[\vspace{6pt}\titlerule\titlerule\titlerule]

\makeatletter
\renewcommand{\section}{\@startsection{section}{0}{0mm}%
{\baselineskip}{.5\baselineskip}%
{ \sffamily\Large\textbf}}%
\makeatother

\makeatletter
\renewcommand{\subsection}{\@startsection{subsection}{1}{2mm}%
{\baselineskip}{.3\baselineskip}%
{\sffamily\Large\textbf}}%
\makeatother

\makeatletter
\renewcommand{\subsubsection}{\@startsection{subsubsection}{2}{4mm}%
{\baselineskip}{.15\baselineskip}%
{\sffamily\large\textbf}}%
\makeatother

\makeatletter
\renewcommand{\paragraph}{\@startsection{paragraph}{3}{6mm}%
{\baselineskip}{.15\baselineskip}%
{\sffamily\large\textbf}}%
\makeatother

\setmarginsrb{2.5cm}{1.5cm}{2.5cm}{2cm}{1cm}{1cm}{1cm}{1cm}

\begin{document}

\begin{center}
\Large{Automatic polishing process of plastic injection molds on a 5-axis
milling center}
\end{center}

\begin{center}
\large{Journal of Materials Processing Technology}
\end{center}

\begin{center}
Xavier Pessoles, Christophe Tournier* 
\end{center}

\begin{center}
 LURPA, ENS Cachan, 61 av du pdt Wilson, 94230 Cachan, France \\
christophe.tournier@lurpa.ens-cachan.fr, Tel: 33 147 402 996, Fax: 33 147 402
211
\end{center}

\subsection*{Abstract}
The plastic injection mold manufacturing process includes polishing operations when surface roughness is critical or mirror effect is required to produce transparent parts. This polishing operation is mainly carried out manually by skilled workers of subcontractor companies. In this paper, we propose an automatic polishing technique on a 5-axis milling center in order to use the same means of production from machining to polishing and reduce the costs. We develop special algorithms to compute 5-axis cutter locations on free-form cavities in order to imitate the skills of the workers. These are based on both filling curves and trochoidal curves. The polishing force is ensured by the compliance of the passive tool itself and set-up by calibration between displacement and force based on a force sensor. The compliance of the tool helps to avoid kinematical error effects on the part during 5-axis tool movements. The effectiveness of the method in terms of the surface roughness quality and the simplicity of implementation is shown through experiments on a  5-axis machining center with a rotary and tilt table.

\subsection*{Keywords}
Automatic Polishing, 5-axis milling center, mirror effect, surface
roughness, Hilbert's curves, trochoidal curves

\begin{tabular}{p{.2\textwidth}p{.7\textwidth}}
 \multicolumn{2}{l}{\textit{Geometric parameters}}\\
 $C_E\left( X_E,Y_E,Z_E\right)$ &  tool extremity point \\
 $\left(u,v \right) $& coordinates in the parametric space \\
 &\\
 \multicolumn{2}{l}{\textit{Trochoidal curve parameters}}\\
 $s$ & curvilinear abscissa \\
 $\mathbf {C}\left(s\right)$ & parametric equation of the guiding curve\\
 $\mathbf {P}\left(s\right)$ & parametric equation of the trochoide curve  \\
 $\mathbf{n}\left(s\right)$  & normal vector of the guiding curve \\
 $p$ & step of the trochoid \\
 $D_{tr}$ & diameter of the usefull circle to construct the trochoid
 \\
 $A$ & amplitude of the trochoid \\
 $Step$ & step between two loops of trochoide\\
\end{tabular}

\begin{tabular}{p{.2\textwidth}p{.7\textwidth}}
\multicolumn{2}{l}{\textit{Technological parameters}}\\
 $D$ & tool diameter \\
 $D_{eff}$ & effective diameter of the tool during polishing \\
 $E$ & amplitude of the envelope of the polishing strip \\   
 $e$ & displacement induced by the compression of the tool\\
 $\theta$ & tilt angle of the tool axis\\
 $\mathbf{u}\left(i,j,k\right)$ & tool axis \\
 $\mathbf{f}$ & tangent vector of the guide curve\\
 $C_c$ & point onto the trochoidal curve \\
\end{tabular}

\begin{tabular}{p{.2\textwidth}p{.7\textwidth}}
\multicolumn{2}{l}{\textit{Machining parameters}}\\
$N$ & spindle speed\\
$V_c$ & cutting speed\\
$V_f$  & feed speed\\
$f_z$ & feed per cutting edge \\
$a_p$ & cutting depth\\
$a_t$ & working engagement\\
$T$ & machining time\\
\end{tabular}

\begin{tabular}{p{.2\textwidth}p{.7\textwidth}}
\multicolumn{2}{l}{\textit{Surface roughness parameters}}\\
$Ra$ & arithmetic average deviation of the surface (2D)\\    
$Sa$ & arithmetical mean height of the surface (3D) \\
$Sq$ & root-mean-square deviation of the surface\\
$Ssk$ & skewness of topography height distribution\\
$Sku$ & kurtosis of topography height distribution\\
\end{tabular}

\newpage

\section{Introduction}
\label{intro}

The development of High Speed Machining (HSM) has dramatically modified the organization of plastic injection molds and tooling manufacturers. HSM in particular has made it possible to reduce mold manufacturing cycle times by replacing spark machining in many cases. In spite of these evolutions, HSM is not enable to remove the polishing operations from the process.
In this paper, we deal with the realization of surfaces with high quality of surface finishing and mirror effect behavior. This means that the part must be perfectly smooth and reflective, without stripes. Such a quality is for example necessary on injection plastic mold cavities in order to obtain perfectly smooth or completely transparent plastic parts.   
From an economic point of view, polishing is a  long and tiresome process requiring much experience. As this process is expensive in terms of price and downtime of the mold, automatic polishing has been developed. Our objective is to use the same means of production from machining to polishing, leading to cost reduction. The aim of the paper is thus to propose a method of automatic polishing on a 5-axis machine tool.

Literature provides various automated polishing experiments. Usually, the
polishing is carried out by an anthropomorphic robot, \cite{Wu07}.
Anthropomorphic robots are used for two main reasons. First, their number of
axes enables them to have an easy access to any area of complex form. Second, it
is possible to attach a great variety of tools and particularly spindles
equipped with polishing force control mechanisms. Automatic polishing studies
have been also carried out on 3 or 5-axis NC milling machine with specially
designed tool to manage polishing force \cite{Huissoon02} as well as on
parallel robots \cite{Roswell06}.
 
Indeed, the polishing force is a key parameter of the process. The abrasion rate
increases when the polishing pressure increases \cite{Lin02}. But as
mentioned in \cite{Roswell06} the contact pressure depends on the polishing
force and also on the geometrical variations of the part. 
An adequate polishing force facilitates the removal of cusps and stripes left on
the part during milling or previous polishing operations. Nevertheless, the
contact stress has to be as constant as possible to avoid over-polishing and
respect form deviation tolerances. Many authors have thus chosen to develop
abrasive systems allowing a dynamic management of the polishing force. In
\cite{Nagata07}, Nagata et al. use an impedance model following force
control to adjust the contact force between the part and the sanding tool. In
\cite{Ryuh06}, Ryuh et al. have developed a passive tool, using a pneumatic
cylinder to provide compliance and constant contact pressure between the surface
and the part. 
A passive mechanism is also used in \cite{Mizugaki90}. The contact force is given by the compressive force of a spring coil.

\begin{center}
  \begin{figure}[!ht]
    \centering
    \includegraphics[width=.8\textwidth]{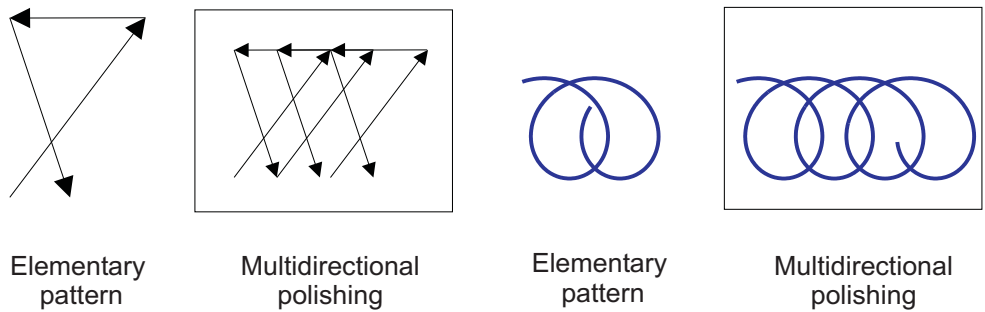}
    \caption{Manual polishing patterns }
    \label{fig:patterns}
  \end{figure}
 \end{center}

In order to carry out an automatic polishing, it is important to use adapted
tool trajectories. According to  \cite{Tsai06}, polishing paths should be
multidirectional rather than monotonic, in order to cover uniformly the mold
surface and to produce fewer undulation errors. Moreover, the multidirectional
polishing path is close to what is made manually. If we observe manual
polishers, we can notice that they go back on surface areas according to various
patterns such as trochoidal polishing paths (or cycloidal weaving paths
\cite{Tsai06} (fig \ref{fig:patterns}). Therefore, it could be profitable
to follow such a process in order to obtain the required part quality. For
instance, some papers use fractal trajectories like the Peano Curve fractal
\cite{Mizugaki92}, which is an example of a space-filling curve, rather
than sweepings along parallel planes \cite{Tam98}. 

This brief review of the literature shows that there is no major difficulty in using a 5-axis machine for automatic polishing with a passive tool. This paper aims at showing the feasibility of automatic polishing using 5-axis machine tools and proposing some polishing strategies.
In the first section, we expose how automatic polishing is possible using a 5-axis HSM center. In particular, we present the characteristics of the passive and flexible tools used. A specific attention is paid to the correlation between the imposed displacement of the tool and the resulting polishing force. Once the feasibility of 5-axis automatic polishing is proved, the various dedicated polishing strategies we have developed are detailed in section 2. These strategies are for the most part issued from previous experiences as for fractal tool trajectories coming from robotized polishing or cycloidal weaving paths representative of manual polishing. In section 3, the efficiency of our approach is tested using various test part surfaces. All the parts are milled then polished on the same production means: a 5-axis Mikron UCP710 milling centre. In the literature, the effectiveness of polishing is evaluated using the arithmetic roughness $Ra$ \cite{Huissoon02}. However, as it is a 2D parameter, this criterion is not really suited to reflect correctly the 3D polished surface quality. We thus suggest qualifying the finish quality of the polished surface through 3D parameters. This point is discussed in the last section as well as the comparison of the surface roughness obtained using automatic polishing with that obtained using manual polishing, a point hardly addressed in the literature. 3D surface roughness measurements are performed using non-contact measuring systems.

\section{Experimental Procedure}

\subsection{Characteristics of the tools}

\begin{center}
  \begin{figure}[!ht]
    \centering
    \includegraphics[width=.7\textwidth]{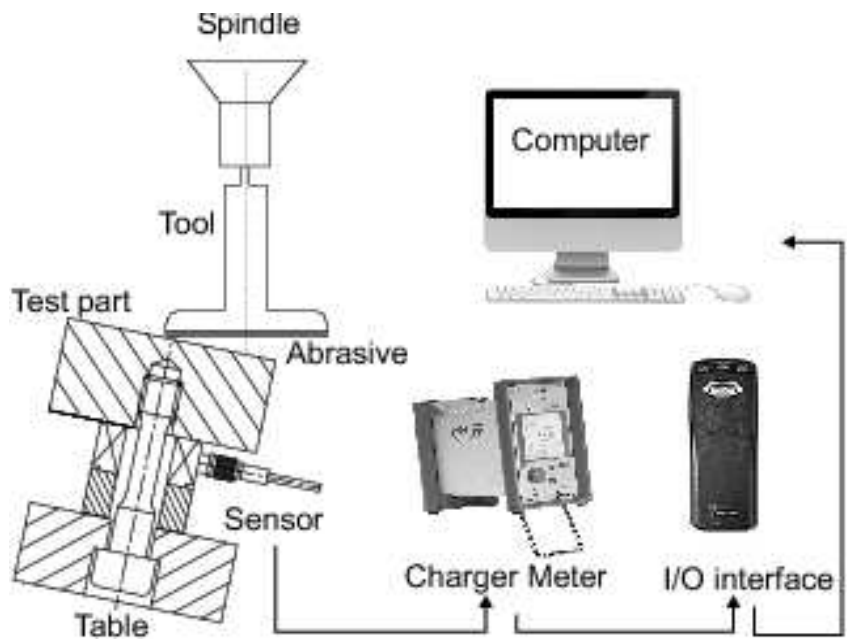}
    \caption{Experimental set-up }
    \label{fig:experience}
  \end{figure}
 \end{center}

As said previously, our purpose is to develop a very simple and profitable system. Therefore, the tools used are the same than those used in manual polishing. The polishing plan is divided into two steps, pre-polishing and finishing polishing. Pre-polishing is performed with abrasive discs mounted on a suitable support. The abrasive particle size is determined by the Federation of European Producers of Abrasives standard (FEPA).
This support is a deformable part made in an elastomer material fixed on a steel
shaft that allows mounting in the spindle. We thus deal with a passive tool.
Hence, we do not have a force feedback control but a position one. We have
studied the relationship between the deflection of the disc support and the
polishing force applied to the part. To establish this relationship, we use a
Quartz force sensor Kistler 9011A mounted on a specially designed part-holder.
The sensor is connected to a charger meter Kistler 5015 itself connected to the
computer through a data-collection device Vernier LabPro to save the data. The
experimental system is depicted in figure \ref{fig:experience}. In addition, the
used sensor  is a dynamic sensor. The effort must therefore change over time
otherwise there would be a drift of the measure. To do so, the movement imposed
on the tool over time is a triangular signal. 

In order to ensure the evacuation of micro chips during the polishing and
guarantee a nonzero abrasion speed at the contact between the part and the tool,
the tool axis $\mathbf{u}$ is tilted relatively to the normal vector to the
polished surface $\mathbf{n}$ and to the feed direction $\mathbf{f}$. The tilt
angle $\theta$ (figure \ref{fig:depincageplan}) is defined as follows:

\begin{center}
  \begin{figure}[!ht]
    \centering
      \includegraphics[width=.4\textwidth]{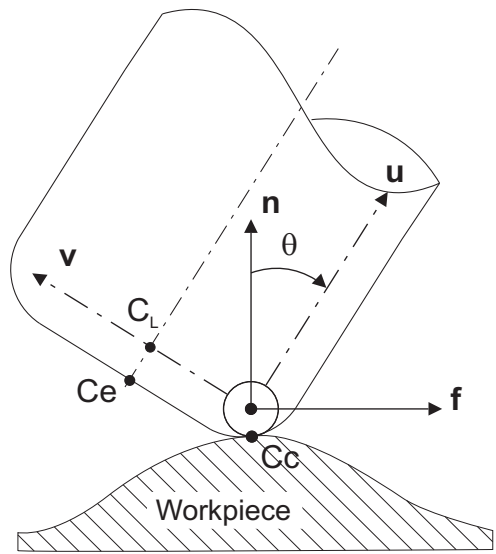}
    \caption{Tool axis tilting}
    \label{fig:depincageplan}
  \end{figure}
\end{center}

\begin{equation}
\label{eq:axeoutil}
\mathbf{u} = \cos\theta \cdot \mathbf{n} + \sin\theta \cdot \mathbf{f} 
 \end{equation}

\begin{center}
  \begin{figure}[!ht]
    \centering
    \includegraphics[width=.7\textwidth]{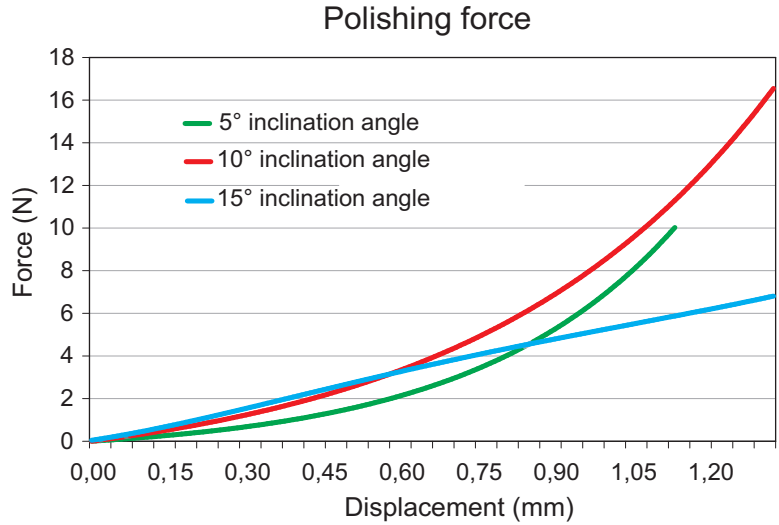}
        \caption{Polishing forces vs displacement}
    \label{fig:forces}
  \end{figure}
 \end{center}

Polishing tests have been conducted considering three different tilt angles
$(5,10,15)$ between the tool axis and the normal vector to the surface in the
feed direction. The correlation between the tool deflection and the polishing
force is shown in figure \ref{fig:forces}. 

The green curve (5 deg) is interrupted because the abrasive disks unstick when the tool deflection is too large. In this configuration, the tilt angle is too low and the body of the disk support, which is more rigid, comes in contact with the workpiece, which deteriorates and unsticks the disk. With a 10 or 15 degrees tilt angle, this phenomenon appears for a higher value of tool deflection, outside the graph.
However, low tilt angle configurations allow faster tool movements since the rotation axes of the 5-axis machine tool are less prompted \cite{Lavernhe08}. Furthermore, it has been showed that trochoidal tool paths require a dynamic machine tool to respect the programmed feedrate \cite{Rauch07}. Then in simultaneous 5-axis configurations, polishing time will be greater with low tilt angles. In addition, the flexibility of the tool will help to reduce or avoid 5-axis kinematic errors \cite{Munlin04}.  Indeed, interfences between the tool and the part could happen because of great tool axis orientation evolutions between two succesive tool positions. Therefore, the disc support deflection would avoid the alteration of the mold surface.

If one considers the law of Preston \cite{Preston27}, the material removal rate $h$ in polishing is proportional to the average pressure of contact, $P$, and to the tool velocity relative to the workpiece, $V$ :
 
 \begin{equation}
   \label{eq:preston}
   h= K_{P} P V
 \end{equation}
   
where $K_{P}$ is a constant ($\frac{m^2 \cdot s}{N}$) including all other parameters (part material, abrasive, lubrification, etc.). Hence, in order to reach an adequate contact pressure, we must increase the tool deflection and consequently we raise the shear stress and the disk unsticks. From a kinematical behavior point of view, low rotational axes movements lead to decrease the polishing time. So we must use a rather low tilt angle (5-10 degrees) and a quite high tool deflection to ensure a satisfactory rate of material removal.

\subsection{5-axis polishing tool path planning}

To generate the polishing tool path, the classical description of the tool path in 5-axis milling with a flat end cutter is used. This leads to define the trajectory of the tool extremity point $C_{E}$ as well as the orientation of the tool axis $\mathbf{u}$ $(i,j,k)$ along the tool path.
With regards to polishing strategy, we use trochoidal tool paths in order to imitate the movements imparted by the workers to the spindle. To avoid marks or specific patterns on the part, we choose to generate trochoidal tool path on fractal curves in order to cover the surface in a multidirectionnal manner. We use more particularly Hilbert's curves which are a special case of the Peano's curve. These curves are used in machining as they have the advantage of covering the entire surface on which they have been generated \cite{Griffits94}. We will develop below the description of the Hilbert's curve which is used as a guide curve for the trochoidal curve then we will examine the trochoidal curve itself.

\subsubsection{Hilbert's curve definition}

The use of fractal trajectories presents two major interests. The first one is that tool paths do not follow specific directions which guarantees an uniform polishing. The second one is linked to the tool path programming. Indeed, tool paths are computed in the parametric space $u,v$ of the surface, that is restricted to the $\left[0,1\right]^2$ interval. Hilbert's curves are known as filling curves, covering the full unit square in the parametric space \cite{Sagan94}, and consequently, the Hilbert's curves fill the 3D surface to be polished.
Hilbert's curves can be defined with a recursive algorithm, the $n$-order curve is defined as follows: 
\begin{itemize}
\item If $n=0$ :
\begin{equation}
\left\{
  \begin{array}{c}
    x_0 = 0\\
    y_0 = 0
  \end{array}
\right.
\end{equation} 
\item Else :
\begin{equation}
\left\{
  \begin{array}{c c c c c c c c}
    x_n &=& 0.5[ & -0.5+y_{n-1}& -0.5+x_{n-1} &0.5+x_{n-1} &0.5-y_{n-1}&]\\
    y_n &=& 0.5[ & -0.5+x_{n-1}& 0.5+y_{n-1} &0.5+y_{n-1}& -0.5-x_{n-1}& ]
  \end{array}
\right.
\end{equation} 
\end{itemize}

\begin{figure}[!ht]
  \centering
    \includegraphics[width=.9 \textwidth]{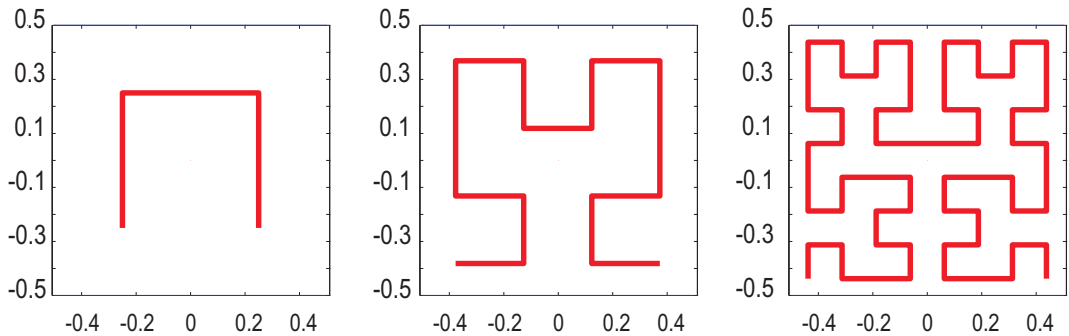}
  \caption{Hilbert's curves (first, second and third order curve)}
  \label{fig:peano123}
\end{figure}

It is then easy to compute first, second or third-order and so Hilbert's curves
(fig \ref{fig:peano123}).

\begin{figure}[!ht]
  \centering
  \includegraphics[width=.5\textwidth]{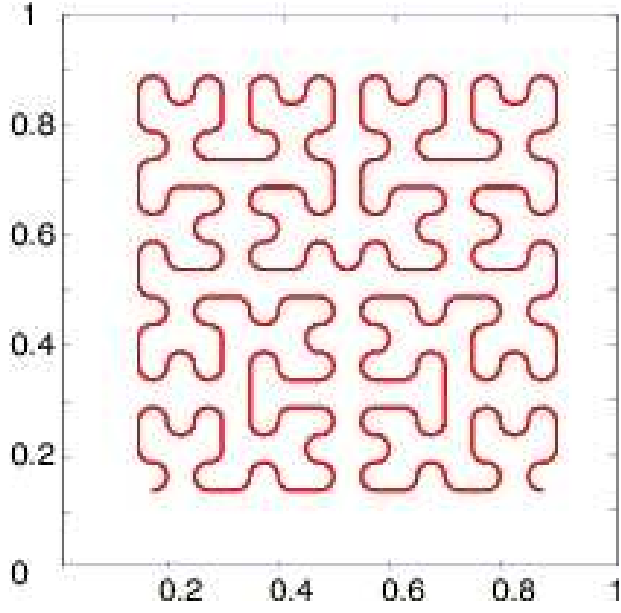}
  \caption{Fourth order cornered Hilbert's curve}
  \label{fig:hilbertc}
\end{figure}

\begin{figure}[!ht]
  \centering
  \includegraphics[width=.65\textwidth]{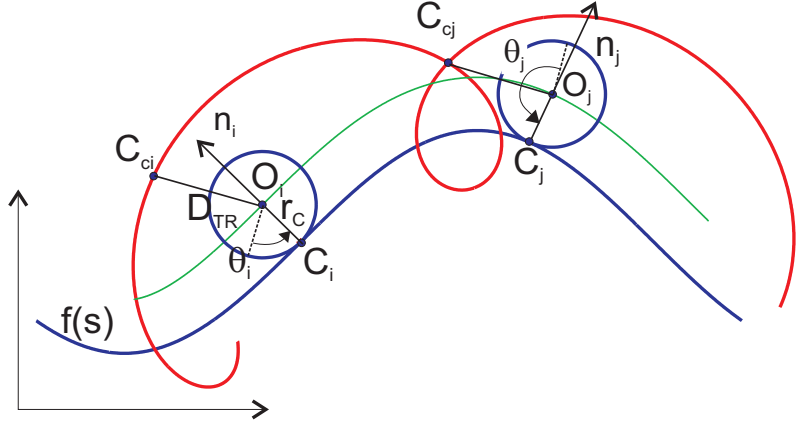}
  \caption{Polishing trajectories on a convex free form} 
  \label{fig:hilbertsurf}
\end{figure}

In order to maintain a tangency continuity along the Hilbert's curve which is
the guide curve of the trochoidal tool path, we have decided to introduce
fillets on the corners of the polishing fractals. Otherwise, at each direction
change on the fractal curve, the polishing tool path would be discontinuous.
Resulting Hilbert's curve is depicted in figure \ref{fig:hilbertc}. Based on
this representation, the curve is easy to manipulate. For example, one could
project this parametric representation directly in the 3D space or use it as the
guide curve for building trochoidal curves (fig \ref{fig:hilbertsurf}) as can be
seen in the next section.

\subsubsection{Mathematical definition of trochoidal curves}

\begin{center}
  \begin{figure}[!ht]
    \centering
      \includegraphics[width=0.6\textwidth]{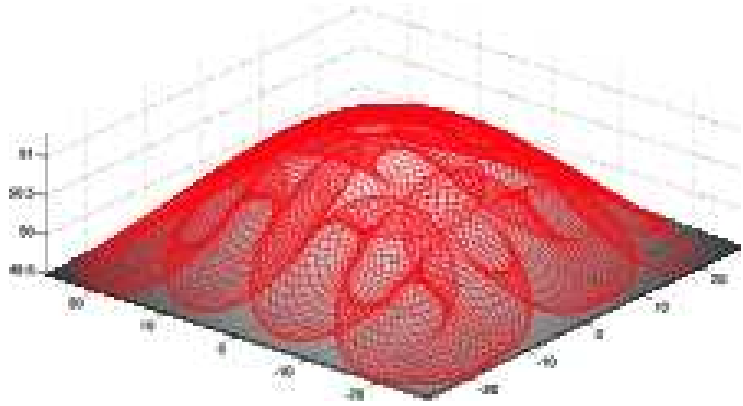}
  \caption{Trochoidal curve parameters}
  \label{fig:DefTrocho}
\end{figure}
\end{center}

Based on the description of trochoidal curves proposed in
\cite{Yates}, we define a trochoidal curve as follows. 
Let $C(s)$ be a 2D parametric curve, where $s$ is the curvilinear length (fig
\ref{fig:DefTrocho}).

$ \mathbf {C}(s) = (s,f(s))$ is the guide curve of the trochoidal curve and $\mathbf {n}(s)$ the normal vector to the curve $C(s)$ at the considered point.
$p$ is the step of the trochoidal curve and we denote $D_{tr}$ its diameter.
The parametric equation of the trochoidal curve is the following: 

\begin{equation}
  \label{eq:trochoidefinale}
  \mathbf {P}(s) = \mathbf {C}(s)
  +\frac{p}{2\pi} \mathbf{n}(s)
  +  D_{tr}
\left(
  \begin{array}{c c}
    \cos(\frac{2\pi s}{p}) & \sin(\frac{2 \pi s}{p})\\
    -\sin(\frac{2\pi s}{p}) & \cos(\frac{2\pi s}{p})
  \end{array}
\right)
\mathbf {n}(s)
\end{equation}

\begin{center}
  \begin{figure}[!h]
    \centering
      \includegraphics[width=0.9\textwidth]{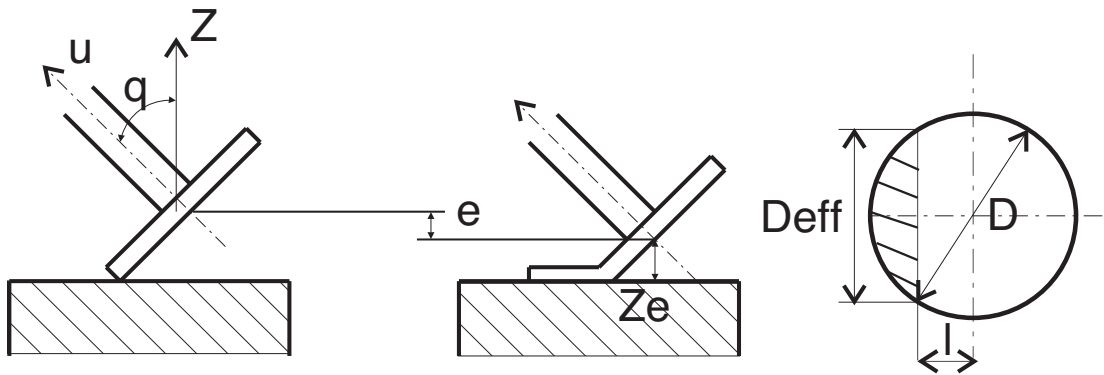}
    \caption{Contact area between the tool and the part}
    \label{fig:ecrasement}
  \end{figure}
\end{center}

The issue is now to link the trochoidal curve parameters to the polishing parameters. 
The amplitude $A$ of the trochoidal curve is equal to twice its diameter
$A=2\cdot D_{tr}$. However, from a  tool path generation point of view, we are
more interested in the tool envelope amplitude than in the trochoidal curve
amplitude. One of the difficulties of modelling the envelope surface of the tool
movement is the tool itself, as abrasive polishing tools are mounted on flexible
supports. The tool polishing amplitude depends on the contact surface between
the tool and the part. This contact is influenced by the tilt angle $\theta$,
the tool diameter $D$ and the imposed tool displacement $e$ to be able to polish
the surface. Indeed, when the tool is laid flat, the contact area is a disk, as
can be seen in figure \ref{fig:ecrasement}. However, when the tool is tilted and
a given displacement $e$ is imposed to the tool, the contact area is a disc
portion.

The effective tool diameter can be computed with the following expressions: 
    
 \begin{equation}
 D_{eff}=2\sqrt{\left(\frac{D}{2}\right)^2- \left(l\right)^2}
 \end{equation}
with: 
 \begin{equation}
l = \frac{\frac{D}{2}\sin \theta -e}{\tan \theta}
 \end{equation}
and: 
 \begin{equation}
E= A+2\frac{D_{eff}}{2}=2D_{tr} + D_{eff}
 \end{equation}

 This yields to the definition of the parameter $D_{tr}$ adjusted to build the trochoidal curve.
 
 \begin{equation}
   \label{eq:regle1}
   D_{tr} = \frac{E-D_{eff}}{2} = \frac{D_{eff}}{6} =
   \frac{1}{3} \sqrt{ \left( \frac{D}{2} \right)^2 -
     \left(\frac{\frac{D}{2}\tan \theta - e}{\tan \theta} \right)^2} 
 \end{equation}

\subsubsection{Tool path generation}

Whatever the nature of the considered surface, the polishing tool paths generation consists of three steps: computation of the tool path in the parametric space, computation of the resulting tool path in the 3D space and computation of the tool axis orientation. Tool path generation relies on the trochoidal curve as described above. The trajectory is defined discretly. The only difficulty is to calculate the normal vector. This is done by using the points $C_{i-1}$ and $C_{i+1}$ and by calculating the next cross product:
 \begin{equation}
 \mathbf{n}_i = \mathbf {Z} \wedge \overrightarrow{C_{i-1} C_{i+1}} 
  \end{equation}

We now describe the method for calculating the direction of the tool axis $\mathbf{u}$ (figure \ref{fig:depincageplan}).
In a first approach we only use the tilt angle defined in the plane $(\mathbf{f};\mathbf {n})$ where $\mathbf f$ is the tangent vector to the guide curve, i.e., the Hilbert's curve and 
$\mathbf {n}$ the normal vector to the machined surface. The tool axis $\mathbf{u}$ is tilted in relation to the Hilbert's curve tangent $\mathbf{f}$ rather than to the trochoidal curve in order to minimize the movements' amplitude of the rotational axes of the machine tool. 

In order to compute the tangent vector  $\mathbf{f_i}$ at the contact point $C_{Ci}$ between the tool and the part, the following expression is used:

 \begin{equation}
  \mathbf {f_i}=\mathbf {n} \wedge 
\underbrace{\left(
\frac{\overrightarrow{C_{Ci}C_{C(i+1)}} \wedge \mathbf {n}}
{||\overrightarrow{C_{Ci}C_{C(i+1)}}||} 
\right)}
 \end{equation}

The location of the tool extremity $C_E$, which is the driven point during machining, depends on the polishing mode, i.e., by pulling or pushing the tool.
The polishing mode is defined by the parameter $\delta$: 

 \begin{equation}
\overrightarrow{OC_E} = \overrightarrow{OC_C} + r \cdot \mathbf {n} +
\delta \left( R-r \right) \cdot \mathbf {v} - r\cdot \mathbf {u}
-e \cdot \mathbf {z}
 \end{equation}
 
 with: 
  \begin{equation}
  \mathbf {v} =  \frac{\mathbf {u} \wedge \mathbf {n}}{||\mathbf {u} \wedge \mathbf {n}||} \wedge \mathbf {u}
  \end{equation}

by noting $\delta=1$ when $\theta > 0$ and $\delta=-1$ when $\theta < 0$.

\subsection{Experiments}
\begin{center}
  \begin{figure}[!h]
    \centering
    \includegraphics[width=0.6\textwidth]{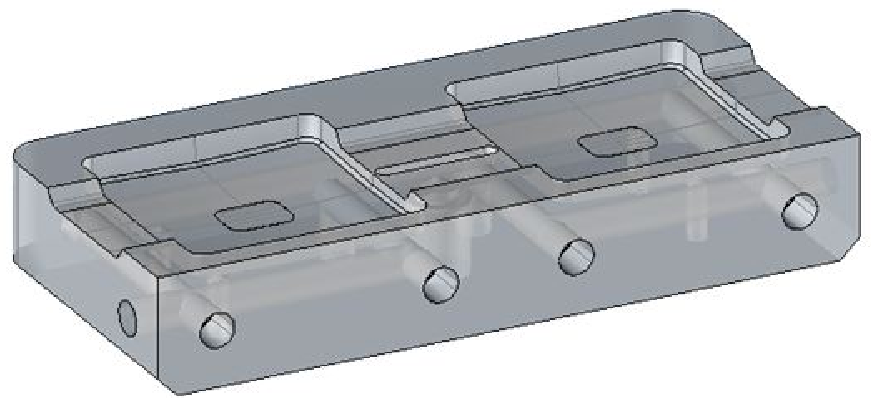}
    \includegraphics[width=0.4\textwidth]{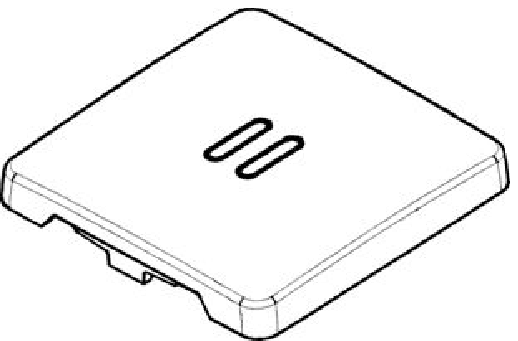}
    \caption{Industrial part and mold}
    \label{fig:devices}
  \end{figure}
 \end{center}

Within the context of plastic mold injection, we study more particularly the
injection molds of electrical devices such as power sockets and switches (figure
\ref{fig:devices}). We thus deal with small slender surfaces. In order to test
our approach, we use two single patch test surfaces, a plane surface and a
convex surface, whose curvature is a little bit larger than the mold curvature.
These are machined in $50x50 mm$ section bloks made of X38CrMoV5 steel. The
Rockwell hardness of the part is 53 HRC after heat treatment.

The part is machined on a 5-axis Mikron UCP710 machine tool to reach a milling
finishing state before polishing. We use four different grades of abrasive for
pre-polishing (FEPA 120, 240, 600, 1200). Abrasives are 18mm diameter disks
sticked on the flexible support.

For 120, 240 and 600 grades, the abrasive is made of aluminium oxide (Al2O3)
within a polymeric material and of silicon carbide on paper for 1200 grade. For
the final polishing, we use three synthetic diamond pastes of grade 9$\mu m$,
3$\mu m$ and 1$\mu m$. Concerning the tool path,  trochoidal trajectories based
on Hilbert's curves for final polishing and based on straight lines for
pre-polishing are used. The milling and polishing sequence is summarized in
table \ref{tab:planning}.

In order to reach high feedrates during polishing, the optimization features of
the Siemens Sinumerik 840D controller have been activated. In particular, the
real-time inverse kinematical transformation (TRAORI) as well as real-time
polynomial interpolation (COMPCURV) to produce smoother axes movements.

\newpage

\section{Results and discussion}

Industrially, the resulting quality of the polished surfaces is first validated by the polisher himself by visual inspection. Surface roughness measurements with contact devices is strictly forbidden to avoid surface damages. However, new non-contact measuring technologies allow the scanning of the 3D topography of the part and data processing according to the international standard for 3D surface roughness.
The International standard \cite{ISO25178} aims at characterizing 3D surface roughness through numerous parameters. Among them, it is important to identify those which are the most appropriate to qualify a mirror effect behavior. To our knowledge, there are no 3D parameter set features of a mirror effect surface. Industrial practices indicate only a $Ra$ around $20nm$. 

A study by  Suh et al. \cite{Suh03} on the surface texture parameters shows that $Sa$ and $Sq$ parameters are not adequate to identify scratches on surfaces. They advise to use the parameter $Ssk$. Hilerio et al. \cite{Hilerio04} also give an interpretation of the standard criteria $Ssk$ and $Sku$ in the context of the control of a polished prosthetic knee. 

$Ssk$ represents the symmetry of the profile:

\begin{itemize}
\item Ssk $=$ 0: profile is symmetrical to the median line,
\item Ssk $>$ 0: profile has more peaks than valleys, 
\item Ssk $<$ 0: profile has more valleys than peaks.
\end{itemize}

$Sku$ represents the distribution averaging:
\begin{itemize}
\item Sku $>$ 3: the distribution is wide (the surface is rather plane),
\item Sku $<$ 3: the distribution is tighted (the surface has a tendency to present peaks or valleys).
\end{itemize}

\begin{center}
  \begin{figure}[!ht]
    \centering
    \includegraphics[width=0.3\textwidth]{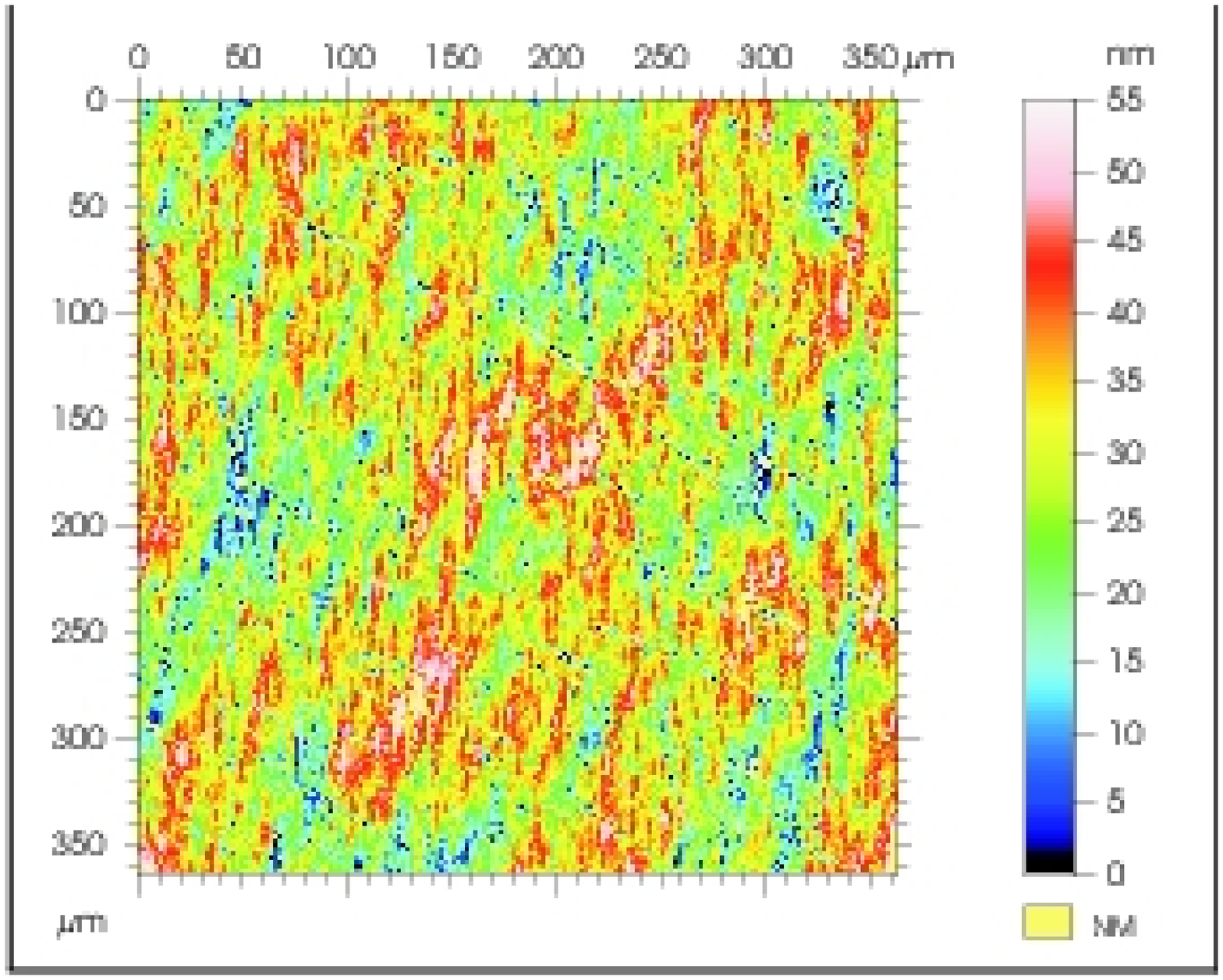}
    \includegraphics[width=0.3\textwidth]{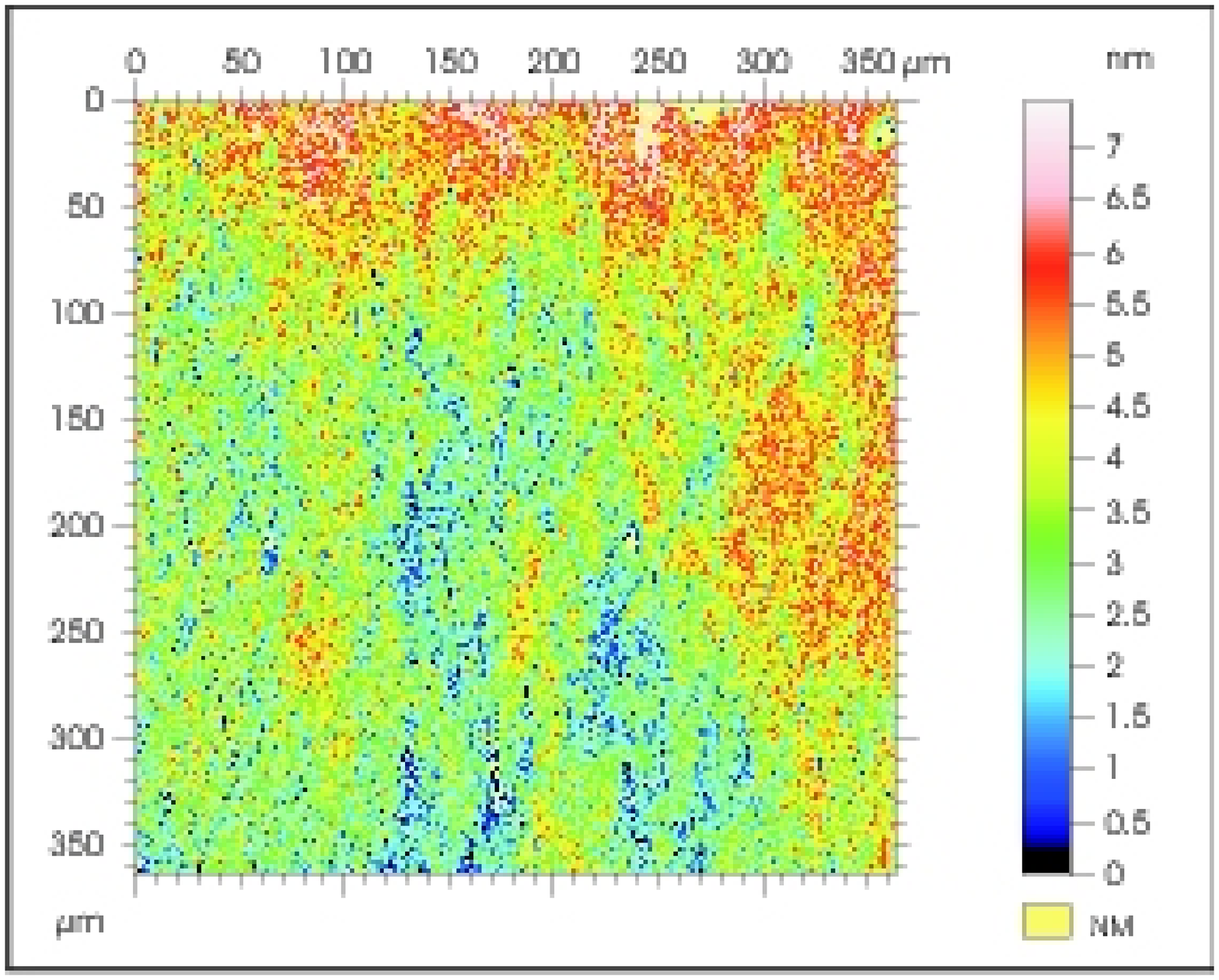}
    \includegraphics[width=0.3\textwidth]{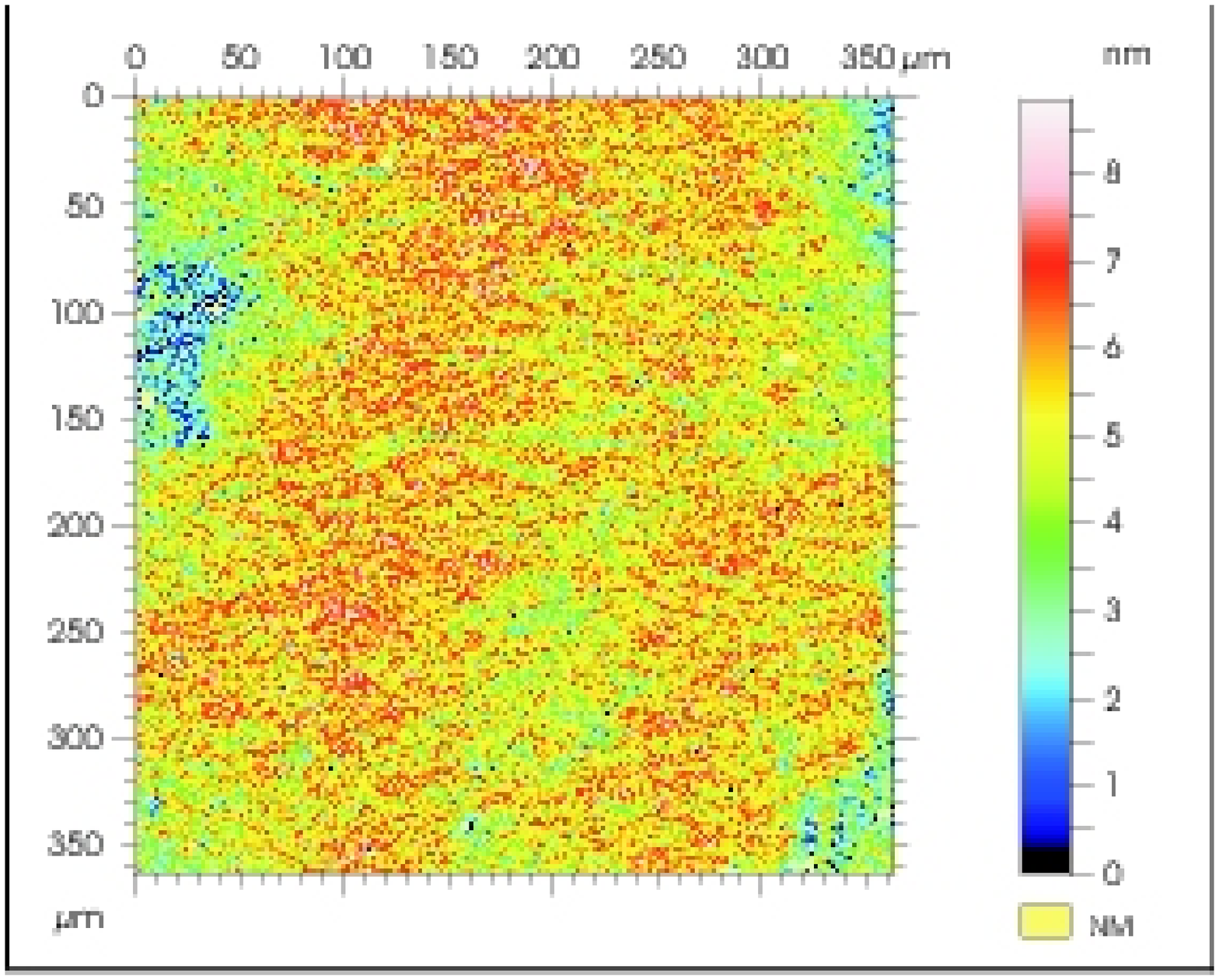}
    \caption{ 3D surface roughness : convex surface (top), planar automatic
(middle), planar manual (bottom)}
    \label{fig:roughness}
  \end{figure}
 \end{center}

Once the parts are polished, we perform 3D surface roughness measurements using a non-contact measuring system (Talysurf CCI 6000). 
We perform measurements on polished parts with our approach (the plane and the
convex surface) and on a plane that has been polished manually by a professional
(figure \ref{fig:roughness}). Measurement results are reported in table
\ref{tab:micro3D}.

\begin{table}[!h]
  \begin{center}
        \begin{tabular}{|c|c|c|c|c|c|c|c|}
                \hline
                 Operations   &N &Vc & $Vf$ & fz & ap & at & T \\ 
                 
                Tool   & $rpm$  & $m/min$ &  $mm/min$ & $mm/tooth$ & $mm$ & $mm$
& min \\ 
                \hline
                parallel planes & 9438& 89 & 8000 & 0,057 & 0,2 & 0,03 & 35  \\
                
                 End mill ($\phi$ 3)  & & & & & & & \\
                \hline
                \hline
                 Operations -- Tools  & $A$ & $Step$ & $Vf$ & $N$ & $\theta$ &
$e$ & $T$ \\ 
                      & $mm$ & $mm$ & $mm/min$ & $rpm$ & $deg$ & $mm$ & $min$\\ 
              \hline
              \hline
          P grade 120 ($\phi$ 18) &12 & 1 &1000 &2000 &3 &0,4 & 15\\
          \hline
          P grade 240 ($\phi$ 18) &12 &1 &1000 &2000 &3 &0,4 & 15\\
          \hline
          P grade 600  ($\phi$ 18) &12 &1 &1000 &2000 &3
          &0,4 & 15\\\hline
          P grade 1200  ($\phi$ 18) &12 &1 &1000 &2000 &3 &0,4 & 15\\\hline
          Diamond abrasive emulsion (9$\mu m$) ($\phi$ 6)  &12 &1 &1000 &2000 &3
&0,3
          & 15\\
          \hline 
         Diamond abrasive emulsion (3$\mu m$)  ($\phi$ 6) &12 &1 &1000 &2000 &3
&0,3
          &  15\\
                    \hline 
          Diamond abrasive emulsion (1$\mu m$) ($\phi$ 6)   &12 &1 &1000 &2000
&3 &0,3
          &  15\\
          \hline 
        \end{tabular}
        \caption{Milling and polishing operations}
   	 \label{tab:planning}
    \end{center} 
     \end{table}
  
   \begin{table}[!h]
   \begin{center} 
    \begin{tabular}{|c|c|c|c|c|}
	\hline
	Surface & Sa & Sq & Ssk & Sku\\ 
	\hline  \hline
	Convex Autom. & 7.619 nm & 9.543 nm & -0.2314 & 2.92 \\ 
	\hline
	Plane Autom. & 1.085 nm & 1.346 nm & 0.103 & 2.713 \\ 
	\hline
	Plane Manual & 1.014 nm & 1.307 nm & -0.5941 & 3.748
	\\ 
	\hline
    \end{tabular}
    \centering
    \caption{3D Roughness parameters}
    \label{tab:micro3D}
    \end{center}
  \end{table}

It can be observed that the convex surface automatically polished presents larger geometric deviations as well as a higher $Sa$ and $Sq$ than those observed for the planar surface. In other words, the rate of material removal is not as good as on the planar surface while trajectories are the same in the $(u,v)$ parametric space. There are several explanations for this behavior.
First, the used polishing pattern, generated in the parametric space, is the same than the planar surface whereas the surface area of the convex part is greater. 	
The result is a lower coverage rate. 	This can also be explained by the machine kinematical behavior during polishing for each part. The planar surface is polished with a 3 axes kinematic while the convex surface requires simultaneous interpolation of the 5 axes of the machine tool. During machining, the relative feedrate between the tool and the part does not match the programmed one due to the slow rotary axes of the 5-axis machining (A: 15rpm; C: 20rpm) \cite{Lavernhe08}. This leads to a slower and less smooth trajectory, reducing the polishing efficiency. 

$Sa$ and $Sq$ are larger for the convex surface but the polished part provides a mirror effect behavior anyway. This confirms the remarks mentioned in \cite{Suh03} and \cite{Hilerio04} as well as the "mean effect" of these parameters.
Mirror effect behavior seems to depend on the $Ssk$ and $Sku$ parameters. Indeed, their values for the convex surface are in adequation with those observed for planar surfaces, which also provide mirror effect behavior.
Regarding the peaks and valleys, the three examples exhibit the same order of magnitude for the parameter $Sku$ with a small advantage for manual polishing. The numerical values correspond fairly well with the observations. 
Finally, we can suppose that the polishing process should be optimized regarding the $Ssk$ and $Sku$
parameters prior to the $Sa$ and $Sq$ parameters.

\section{Conclusion}

In this article, we propose a solution to produce mirror effect polished surfaces on a 5-axis machine tool normally dedicated to molds' milling.
The passive tool used is simple to implement. A preliminary calibration allows us to correlate the polishing force and the tool deviation. We have also developed polishing tool paths similar to the patterns used in manual polishing in order to avoid marks on the polished part. The polishing quality is comparable to the manual method and polishing time is similar. However, in order to maintain a constant coverage rate, we should take into consideration the effective area of the part to be polished when generating the tool paths in the $(u,v)$ plane. 
From roughness point of view, the polishing process must reduce the amplitude of the peaks and valleys which is characterized by a $Sku$ parameter superior to 3.
For complex shapes, kinematic performances of the machine tool are very important in order to guarantee a tool feedrate as high as possible and smooth, thus leading to a good rate of material removal. In particular, rotary axes have to be very performing to respect the programmed feedrate. We seek now to develop polishing strategies that allow the polishing of small radius blending surfaces on the injection moulds. 

\newpage

\section*{Appendix: 3D surface roughness parameters}

$Sa$: Arithmetical mean height of the surface.
  \begin{equation}
  Sa = \frac{1}{A} \iint Z(x,y)  dxdy
  \end{equation}

$Sq$: Root-mean-square deviation of the surface.
This is a dispersion parameter defined as the root mean square value of the surface departures within the sampling area.
  \begin{equation}
  Sq = \sqrt {\frac{1}{A}\iint\limits_A\!Z^2(x,y)dxdy}
  \end{equation}
  
$Ssk$: Skewness of topography height distribution.
This is the measure of asymmetry of surface deviations about the mean plane.
  \begin{equation}
  Ssk = \frac{1}{(Sq)^3}\left[ \frac{1}{A}\iint\limits_A Z^3(x,y) dx  dy\right]
  \end{equation}

$Sku$: Kurtosis of topography height distribution.
This is a measure of the peakedness or sharpness of the surface height distribution.
  \begin{equation}
  Sku = \frac{1}{(Sq)^4}\left[ \frac{1}{A}\iint\limits_A Z^4(x,y) dx dy\right]
   \end{equation}

\newpage

\bibliographystyle{unsrt}

\end{document}